\documentclass[onecolumn,showpacs,preprintnumbers,showkeys]{revtex4}
\usepackage{amsfonts}
\usepackage{amsmath}
\usepackage{graphicx}
\usepackage{dcolumn}
\usepackage{bm}
\usepackage{mathrsfs}
\usepackage{epsfig}

\begin{document}

\title{Tracer-particle dynamics in MHD fluids}
\author{Massimo Tessarotto}
\affiliation{Department of Mathematics and Informatics, Trieste University, Trieste, Italy}
\altaffiliation[Also at ]{Consortium for Magnetofluid Dynamics, Trieste University, Trieste, Italy}
\email{maxtextss@gmail.com}
\author{Claudio Asci}
\affiliation{Department of Mathematics and Informatics, Trieste University, Trieste, Italy}
\altaffiliation[Also at ]{Consortium for Magnetofluid Dynamics, Trieste University, Trieste, Italy}
\author{Claudio Cremaschini}
\affiliation{International School for Advanced Studies (SISSA) and INFN, Trieste, Italy}
\altaffiliation[Also at ]{Consortium for Magnetofluid Dynamics, Trieste University, Trieste, Italy}
\author{Alessandro Soranzo}
\affiliation{Department of Mathematics and Informatics, Trieste University, Trieste, Italy}
\altaffiliation[Also at ]{Consortium for Magnetofluid Dynamics, Trieste University, Trieste, Italy}
\author{Marco Tessarotto}
\altaffiliation[Also at ]{Consortium for Magnetofluid Dynamics, Trieste University, Trieste, Italy}
\affiliation{Civil Protection Agency, Regione Friuli Venezia-Giulia, Palmanova (Udine),
Italy}
\author{Gino Tironi}
\altaffiliation[Also at ]{Consortium for Magnetofluid Dynamics, Trieste University, Trieste, Italy}
\affiliation{Department of Mathematics and Informatics, Trieste University, Trieste, Italy}
\date{\today }

\begin{abstract}
A key issue in fluid dynamics is the unique definition of the phase-space
Lagrangian dynamics characterizing prescribed ideal fluids (i.e., continua),
which is related to the dynamics of so-called \textit{ideal tracer particles}
(ITP) moving in the same fluids. These are by definition particles of
infinitesimal size which do not produce significant perturbations of the
fluid fields and do not interact among themselves. For Navier-Stokes (NS)
fluids, the discovery by Tessarotto et al. (2005-2009) of the phase-space
dynamical system advancing in time the state of the fluid, has made
possible, \textit{in the case NS fluids}, the actual definition of these
trajectories. In this paper we intend to pose the problem in the case of
compressible/incompressible magnetofluids based on the inverse kinetic
theory which can be developed for their phase-space statistical description
(see also accompanying paper) \ We propose the conjecture of the existence
of a subset of ITP's (i.e., particular solutions of the phase-space
dynamical system), denoted as \textit{thermal ideal tracer particles}
(TITP). These particles are characterized by a relative velocity with
respect to the fluid, whose magnitude is determined, by the kinetic pressure
(in turn, related to the fluid pressure).
\end{abstract}

\maketitle



\section{Introduction}

In this paper we report the discovery of a subset of so-called \emph{ideal
tracer particles }(ITP \cite{Tessarotto2009})\emph{\ }belonging to
Navier-Stokes (NS) fluids which are denoted as \emph{thermal tracer
particles }(TTP). Their states are found to be uniquely dependent on the
local state of the fluid\footnote{%
This means that a suitable statistical ensemble of TTPs should reproduce
exactly the dynamics of the fluid. In other words, it should be possible to
determine the fluid fields characterizing the fluid state by means of
suitable statistical averages on the ensemble of TTPs so that they satisfy
\emph{identically} the required set of fluid equations.}. The result applies
to NS fluids described as mesoscopic, i.e., continuous fluids, which can be
either viscous or inviscid, compressible or incompressible, thermal or
isothermal, isentropic or non-isentropic. We shall assume that the state of
these fluids is represented by an ensemble of observables $\left\{ Z(\mathbf{%
r},t)\right\} $ $\equiv \left\{ Z_{i}(\mathbf{r},t),i=1,..,n\right\} $ (with
$n$ and integer $\geq 1$), i.e., \emph{fluid fields,} which can be \emph{%
unambiguously prescribed }as continuous and suitably smooth functions,
respectively, in $\overline{\Omega }\times I$ and in the open set $\Omega
\times I.$ We intend to show that, as a remarkable consequence, the
phase-space dynamical\ system which advances in time (the states of) these
particles can be \emph{uniquely} prescribed in such a way to determines
\emph{self-consistently} the time evolution of the complete set of fluid
equations characterizing the fluid. This implies that TTPs must reproduce
\emph{exactly} the dynamics of the fluid. In other words, by means of an
appropriate statistical averages on the ensemble of the TTPs, it is possible
to determine the time-evolution of the fluid state, in such a way that it
\emph{satisfies identically} the required set of fluid equations.

\subsection{Lagrangian dynamics of ideal tracer particles}

A key aspect of fluid dynamics is the proper definition of the \emph{%
phase-space Lagrangian dynamics} for continuous fluid systems, whereby
possibly \emph{all the fluid fields} characterizing the actual fluid state $%
\left\{ Z(\mathbf{r},t)\right\} $ can be identified with suitable
statistical averages on appropriate ensembles of (fictitious) particles%
\footnote{%
Thus, for example, in the case of an incompressible NS fluid, this would
require to represent both the fluid velocity $\mathbf{V}(\mathbf{r},t)$ and
the fluid pressure $p\mathbf{(r},t)$ in terms of suitable statistical
averages of an appropriate probability density. This goal can be realized by
means of the inverse kinetic theory (IKT) developed in Refs.\cite%
{Tessarotto2004,Ellero2005}}. This refers, in particular, to the phase-space%
\emph{\ }dynamics of so-called \emph{ideal tracer particles} (ITPs \cite%
{Tessarotto2009}), namely rigid\textbf{\ }extended classical particles
immersed in the fluid, all having the same support and infinitesimal size
such that during their motion they \emph{do} \emph{not mutually interact}
and \emph{do not perturb} \emph{the state of the fluid}. Depending on their
inertial mass $m_{P}$, ITPs can belong to different species of particles;
thus, in general, their mass can differ from that of the corresponding
displaced fluid element $m_{F}.$ On the other hand, ITPs carrying the mass $%
m_{P}\equiv m_{F}$ will be denoted as the \emph{NS ideal tracer particles} (%
\emph{NS-ITPs})\emph{. }In the following, in order to characterize the
Lagrangian dynamics of NS fluids, ITPs will be identified only with NS-ITPs.
In this framework, it follows that ITPs can undergo, by assumption, solely
\textquotedblleft unary\textquotedblright\ interactions with external
force-fields and with the continuum fluid. Namely, in both cases they are
subject only to the action of a continuum \emph{mean-field} \emph{%
acceleration}$,$which can depend solely on the states of the particle and of
the fluid, i.e., is of the form $\mathbf{F=F}(\mathbf{x},t)$. As a
consequence, ITPs can be treated as Newtonian point-like particles
characterized by a Newtonian state $\mathbf{x}=(\mathbf{r,v})$ spanning the
phase-space $\Gamma \equiv \Omega \times U,$ with the position $\mathbf{r}$
and the \emph{kinetic velocity} $\mathbf{v}$ belonging respectively to the
configuration space of the fluid $\Omega $ (in the following to be
identified with a bounded subset of $%
\mathbb{R}
^{3}$) and the velocity space $U\equiv
\mathbb{R}
^{3}$.

\subsection{The Navier-Stokes dynamical system}

It is possible to prove \cite{Tessarotto2010} that the state $\mathbf{x}$ of
a generic ITP advances in time by means of a Newtonian classical dynamical
system (DS). Contrary to a widespread misconception such a DS is \emph{%
finite-dimensional}, i.e., it is characterized by a \emph{finite degree of
freedom}. In fact, the DS can be prescribed in terms of a, \emph{generally
non-unique}, vector field of the form $\mathbf{X}(\mathbf{x},t)\equiv
\left\{ \mathbf{v,}\frac{1}{m_{P}}\mathbf{\mathbf{K}}\equiv \mathbf{\mathbf{F%
}}\right\} ,$ with $\frac{1}{m_{P}}\mathbf{K}\equiv \mathbf{F}$ a suitable
mean-field acceleration. This is identified with the flow ($T_{t_{o},t}$)
generated by the initial value problem associated to the deterministic
equations of motion \ (\emph{Newton's equations})%
\begin{equation}
\left\{
\begin{array}{c}
\frac{d}{dt}\mathbf{x}=\mathbf{X}(\mathbf{x},t), \\
\mathbf{x}(t_{o})=\mathbf{x}_{o}.%
\end{array}%
\right.  \label{eqq-1}
\end{equation}%
Such flow is referred to as as \emph{Navier-Stokes dynamical system}
(NS--DS) and is a homeomorphism in $\Gamma $ with existence domain $\Gamma
\times I,$ of the type
\begin{equation}
T_{t_{o},t}:\mathbf{x}_{o}\rightarrow \mathbf{x}(t)=T_{t_{o},t}\mathbf{x}%
_{o},  \label{DYN-S}
\end{equation}%
with $t\in I\subseteq
\mathbb{R}
$, $T_{t_{o},t}$ being a measure-preserving evolution operator associated to
$\mathbf{X}(\mathbf{x},t)$. \emph{\ Thus, by definition, the NS-DS is
uniquely prescribed by the couple }$\left\{ \mathbf{x},\mathbf{X}(\mathbf{x}%
,t)\right\} ,$ with $\mathbf{x}(t)\equiv \mathbf{x}=(\mathbf{r,v})$ to be
identified with the instantaneous state of a generic ITP.

\section{Gedanken experiment}

For a prescribed continuous fluid system, such as a
compressible/incompressible NS thermofluid, the problem arises whether there
might exist a subset of the ensemble of ideal tracer particles, i.e., of the
dynamical system (\ref{DYN-S}), such that their Newtonian state and
corresponding time evolution depend only on the state of the fluid $\left\{
Z\right\} .$

We argue that it should be possible to prove the existence of the TTPs by
performing a conceptual experiment (\emph{Gedanken experiment}) on the
fluid, i.e., looking at the properties of the IKT-statistical models $%
\left\{ f,\Gamma \right\} $. The conjecture is suggested by the following
arguments:

\begin{itemize}
\item The state of the fluid is solely dependent of the fluid fields, which
in the case of a compressible NS thermofluid can be identified with the set $%
\left\{ Z_{1}\right\} .$

\item The time-evolution of $\left\{ Z\right\} $ is necessarily independent
of the KDF $f(\mathbf{x},t)$ and of the NS-DS (\ref{DYN-S}).

\item On the other hand, the time evolution of the fluid fields is also
generated by the Lagrangian IKE in terms of the NS-DS (\ref{DYN-S}).
\end{itemize}

Here we conjecture that the ITPs belonging to such a subset\ should fulfill
the following properties:

\begin{enumerate}
\item \emph{GDE-requirement \#1:} their time evolution should be, at all
times $t\in I,$ independent of the particular form of the KDF $f(\mathbf{x}%
,t).$ As a consequence, for them the form of the mean-field force $\mathbf{F}
$ should be \emph{independent} of the KDF $f(\mathbf{x},t)$ [introduced in
the IKT-statistical model $\left\{ f,\Gamma \right\} $]$;$

\item \emph{GDE-requirement \#2:} for prescribed initial conditions, their
Newtonian states $\mathbf{x}(t)\equiv \mathbf{x}=(\mathbf{r,v}),$ and
equivalently also $\mathbf{y}(t)\equiv \mathbf{y}=(\mathbf{r,u}),$ should
depend solely on the fluid fields $\left\{ Z_{1}\right\} $.

In addition, one should expect that for all TTPs:

\item \emph{GDE-requirement \#3 - Local magnitude of }$\mathbf{u}(t)$: the
magnitude of their instantaneous relative velocity $\left\vert \mathbf{u}%
(t)\right\vert \equiv \left\vert \mathbf{u}(\mathbf{r},t)\right\vert $
remains\ at all times $t\in I$ proportional to the local thermal velocity $%
v_{th}(\mathbf{r},t),$ i.e., of the form
\begin{equation}
\left\vert \mathbf{u}(t)\right\vert =\beta v_{th}(\mathbf{r},t),
\label{CONSTRAINT}
\end{equation}%
with $p_{1}(\mathbf{r},t)>0,\mathbf{r\equiv r}(t)$ and $\beta $ denoting
respectively the \emph{kinetic pressure} (see accompanying paper \cite%
{Tessarotto2011}), the instantaneous position of the same particle and an
appropriate non-vanishing constant, i.e., a function independent of $(%
\mathbf{r},t)$;

\item \emph{GDE-requirement \#4 - Kinetic constraint on the local direction
of }$\mathbf{u}(t)$: Let us introduce for $\mathbf{u}(t)$ the representation
\begin{equation}
\mathbf{u}(t)=\beta v_{th}(\mathbf{r},t)\mathbf{n}(\mathbf{r},t),
\label{REPRESENTATION of u}
\end{equation}%
with $\mathbf{n}(\mathbf{r},t)$ the unit vector proscribing the local
direction of $\mathbf{u}(t).$ Then $\mathbf{n}(\mathbf{r},t)$ should satisfy
the kinetic constraint:
\begin{equation}
\mathbf{n}(\mathbf{r},t)\cdot \nabla \widehat{p}_{1}(\mathbf{r},t)=0.
\label{TANGENCY CONDITION-2}
\end{equation}%
In fact, for a non-uniform kinetic pressure satisfying locally
\begin{equation}
\nabla \widehat{p}_{1}(\mathbf{r},t)\neq 0,  \label{NON-UNIFORM
PRESSURE}
\end{equation}%
the requirement (\ref{CONSTRAINT}) cannot generally be met unless the
relative velocity $\mathbf{u}(t)$ remains tangent to the \emph{local
isobaric surface} $\widehat{p}_{1}(\mathbf{r},t)=const.$ As a consequence,
the direction of $\mathbf{u}(t)$ is necessarily uniquely determined, once
the initial conditions (\ref{eqq-1}) and consequently its initial direction
\begin{equation}
\mathbf{n}(\mathbf{r},t_{o})\equiv \mathbf{n}_{o}(\mathbf{r})
\label{INITIAL DIRECTION}
\end{equation}%
have been set. Hence, in validity of (\ref{NON-UNIFORM PRESSURE}) the unit
vector $\mathbf{n}(\mathbf{r},t)$ must be orthogonal to the unit vector
\emph{\ }
\begin{equation}
\mathbf{b}(\mathbf{r},t)=\frac{\nabla \widehat{p}_{1}(\mathbf{r},t)}{%
\left\vert \nabla \widehat{p}_{1}(\mathbf{r},t)\right\vert },  \label{Omega2}
\end{equation}%
i.e., the \emph{kinetic constraint}
\begin{equation}
\mathbf{n}(\mathbf{r},t)\cdot \mathbf{b}(\mathbf{r},t)=0
\label{TANGENCY CONDITION}
\end{equation}%
must hold identically for all $\left( \mathbf{r,}t\right) \in \Omega \times
I.$

\item \emph{GDE-requirement \#5 - Time evolution of }$\mathbf{n}(\mathbf{r}%
,t):$ \ the unit vector $\mathbf{n}(\mathbf{r},t)$ satisfies an
initial-value problem of the form%
\begin{equation}
\left\{
\begin{array}{c}
\frac{d\mathbf{n}(\mathbf{r},t)}{dt}=\mathbf{\Omega }(\mathbf{r},t)\times
\mathbf{n}(\mathbf{r},t), \\
\mathbf{n}(\mathbf{r}(t_{o}),t_{o})=\mathbf{n}(\mathbf{r}_{o},t_{o}),%
\end{array}%
\right.   \label{EVOLUTION EQUATION for n}
\end{equation}%
with $\mathbf{\Omega }(\mathbf{r},t)$ denoting a suitable pseudo-vector.
Without loss of generality we shall require that: 1) $\mathbf{\Omega }(%
\mathbf{r},t)$ is a smooth real vector function defined in $\overline{\Omega
}\times I;$ 2) $\mathbf{\Omega }(\mathbf{r},t)$ is defined also in the limit
$p_{1}(\mathbf{r},t)\rightarrow 0^{+}.$
\end{enumerate}

An interesting issue concerns the physical interpretation TTP dynamics, and
in particular the evolution equation for the unit vector $\mathbf{n}(\mathbf{%
r},t)$ [i.e., the direction of the particle relative velocity; see figure 1
and Eqs.(\ref{EVOLUTION EQUATION for n})] and related the pseudovector%
\textbf{\ }$\mathbf{\Omega }(\mathbf{r},t)$. \ In fact, that Eqs.(\ref%
{EVOLUTION EQUATION for n}) are analogous to those of a rigid body rotating
with angular velocity $\mathbf{\Omega }_{\mathbf{r}}=-\mathbf{\Omega }(%
\mathbf{r},t)\mathbf{.}$ This suggests that $\mathbf{\Omega }(\mathbf{r},t)$
should be related to the local fluid vorticity $\mathbf{\xi }\equiv \nabla
\times \mathbf{V}$. \ Indeed denoting by $\frac{D\mathbf{b}(\mathbf{r},t)}{Dt%
}\equiv \frac{\partial }{\partial t}+\mathbf{V}(\mathbf{r},t)\cdot \nabla $
the fluid convective derivative, it follows%
\begin{equation}
\mathbf{\Omega }(\mathbf{r},t)\mathbf{\equiv b}(\mathbf{r},t)\times \frac{d%
\mathbf{b}(\mathbf{r},t)}{dt}=\mathbf{b}(\mathbf{r},t)\times \frac{D\mathbf{b%
}(\mathbf{r},t)}{Dt}+\mathbf{b}(\mathbf{r},t)\times \left( \mathbf{u\cdot
\nabla }\right) \mathbf{b}(\mathbf{r},t)  \label{AAA-1}
\end{equation}%
where%
\begin{eqnarray}
&&\left. \mathbf{b}(\mathbf{r},t)\times \left( \mathbf{u\cdot \nabla }%
\right) \mathbf{b}(\mathbf{r},t)=\right.   \label{AAA-2} \\
&&\left. =-\mathbf{\xi \cdot }\left[ \underline{\underline{\mathbf{1}}}%
\mathbf{-bb}\right] +\frac{1}{\left\vert \nabla p_{1}\right\vert }\left[
\mathbf{b\times \nabla }(\nabla p_{1}\cdot \mathbf{V})-\mathbf{b\times }%
\left( \nabla p_{1}\mathbf{\cdot \nabla }\right) \mathbf{V}\right] \right. .
\end{eqnarray}%
In the last equation the first term on the r.h.s. of denotes the tangential
component of the vorticity (i.e., the component belonging to the local
tangential plane w.r. to a local isobaric surface $\widehat{p}_{1}=const.$).
This means that near a vortex the motion of TTPs is qualitatively similar to
that of a rotating rigid body. However, by inspection of the remaining terms
in Eqs.(\ref{AAA-1}) and (\ref{AAA-2}), it is evident that more complex
particle-acceleration effects may be present, which are driven by
time-dependent pressure and velocity-gradients contributions.

\section{Conclusions}

A fundamental issue for Navier-Stokes fluids, is their characterization in
terms of the dynamics of ideal tracer particles (ITPs). Based on the
formulation of an inverse kinetic theory for compressible/incompressible NS
thermofluids (THM.1), in this paper the existence of a subset of ITPs
denoted as TTPs whose states depend solely on the state of the fluid has
been conjectured. A detailed proof of the statement will be reported
elsewhere \cite{Tessarotto2011a}.

Applications of the present theory are in principle several, and deal, in
particular, with the dynamics of small particles (such as solid particles or
droplets, commonly found in natural\ phenomena and industrial applications)
in compressible/incompressible thermofluids. The accurate description of
particle dynamics, as they are pushed along erratic trajectories by
fluctuations of the fluid fields, is essential, for example, in combustion
processes, in the industrial production of nanoparticles as well as in
atmospheric pollutant transport, cloud formation and air-quality monitoring
of the atmosphere.

\section{Acknowledgments}

Work developed in cooperation with the CMFD Team, Consortium for
Magnetofluid Dynamics (Trieste University, Trieste, Italy). \ Research
developed in the framework of the CMFD research programs (Consorzio di
Magnetofluidodinamica, University of Trieste, Italy) and the GDRE (Groupe
des Recherches Europ\'{e}enne) GAMAS, C.N.R.S., France. The support of the
GNFM (National Group of Mathematical Physics) of INDAM (Italian National
Institute for Advanced Mathematics) is acknowledged.

\end{document}